# Deterministic Sub-wavelength Control of Light Confinement in Nanostructures

Giorgio Volpe[1], Gabriel Molina-Terriza[1,2] and Romain Quidant[1,2]

*[1] ICFO-Institut de Ciencies Fotoniques, Mediterranean Technology Park, 08860 Castelldefels, Barcelona, Spain,*

*[2] ICREA-Institució Catalana de Recerca i Estudis Avançats, 08010 Barcelona, Spain*

(Dated: January 11, 2010)

## Abstract

We propose a novel deterministic protocol, based on continuous light flows, that enables us to control the concentration of light in generic plasmonic nanostructures. Based on an exact inversion of the response tensor of the nanosystem, the so-called Deterministic Optical Inversion (DOPTI) protocol provides a physical solution for the incident field leading to a desired near field pattern, expressed in the form of a coherent superposition of high order beams. We demonstrate the high degree of control achieved on complex plasmonic architectures and quantify its efficiency and accuracy.

High concentration of the electromagnetic field, well beyond the diffraction limit, can be achieved in the vicinity of plasmonic nanostructures by creating strong surface charge gradients. The so-called *hot-spots* can for instance be managed in the nano-gap between two adjacent metallic nanoparticles [1] as in optical nano-antennas [2] or at the apex of a sharp metallic tip [3]. The optical near field distribution resulting from the interaction of **light** with plasmonic nanostructures is mostly determined by the geometry of the metal system and the illumination conditions (wavelength, polarization, etc…). For a given geometry, a standard illumination, by a plane wave or a Gaussian beam featuring homogeneous polarization and phase over time and space, only offers poor control of its near field response. While a modification of wavelength or polarization can induce substantial changes, it does not allow one to control the location of a *hot-spot* at any desired position of the nanostructure. Nonetheless, the accurate and dynamical control of the optical near field at the sub-$\lambda$ scale is required for the development of future nano-optical devices. Different strategies have been proposed to reach this goal. A first family of approaches, inspired by coherent control, relies on temporally shaping the phase and amplitude of ultra-short laser pulses illuminating the nanostructures [4-5]. By combining pulse shaping with a learning algorithm, the feasibility of generating user-specified optical near field responses, as first proposed by Stockman et al. [4], was demonstrated experimentally on a star-like silver object [5]. Experimental control of the local optical response of a metal surface was also achieved by adjusting the temporal phase between two unshaped ultrashort pulses [6]. Alternatively, the idea of time reversal has recently been proposed [7].

A different approach, based on continuous wave flows, consists in exploiting the spatial phase shaping (in opposition to temporal phase shaping) of high-order laser beams. We have recently demonstrated how the phase jumps at the focus of Hermite-Gaussian beams enables coupling to plasmonic dark modes in coupled plasmonic antennas. Spatial phase shaping of a Gaussian beam into a higher-order Hermite-Gaussian beam becomes then a way of switching ON and OFF hot-spots [8]. However, achieving a deterministic control of light confinement at any specific location of a generic plasmonic nanostructure requires a procedure that enables us to determine the actual incident field that leads to the desired near field pattern.

Here we propose a novel, fully deterministic numerical protocol towards the control of the optical near-field response of a generic nanosystem in an accurate and dynamical way. The so-called Deterministic Optical Inversion (DOPTI) algorithm enables us to reconstruct, from a desired near field pattern, the closest physical solution for the incident field, expressed on a **basis of focused Hermite-Gaussian (HG) beams**.

In order to draw the general principle of the DOPTI algorithm, let us consider the configuration in which an incident field $\mathbf{E}_{in}$, **after focusing through an objective lens,** impinges on a generic nano-object of dielectric function $\varepsilon_{ob}(\omega)$. Upon linear interaction, the electric field $\mathbf{E}_{out}$ inside the object connects to $\mathbf{E}_{in}$ by:

$$\mathbf{E}_{out} = \mathbb{X}\mathbf{E}_{in}. \tag{1}$$

In this functional equation, $\mathbb{X}$ is a linear operator describing the optical response of the object as derived from Maxwell's equations. A full control of the optical near field response would consist in retrieving the incident field $\mathbf{E}_{in}^{i}$ that leads to the desired output field $\mathbf{E}_{out}^{i}$ (see Fig. 1) by solving:

$$\mathbf{E}_{in}^{i} = \mathbb{X}^{-1}\mathbf{E}_{out}^{i}, \tag{2}$$

**where we assume that $\mathbb{X}$ is invertible. This is usually the case although special attention must be paid when $\mathbb{X}$ is singular [9].**

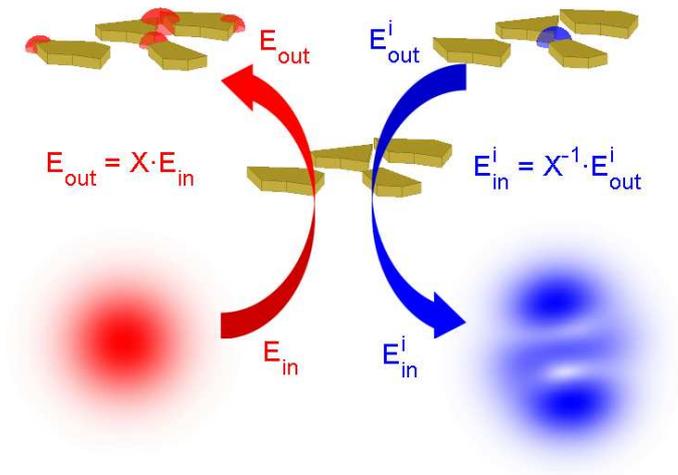

FIG. 1: (color online). Schematic description of the DOPTI algorithm. The left path (red) sketches the situation in which a Gaussian beam ($\mathbf{E}_{\text{in}}$) impinges on the structure leading to the output field ($\mathbf{E}_{\text{out}}$). The right path (blue) sketches the reverse situation in which the optimal incident field $\mathbf{E}_{\text{in}}^{i}$ is computed from the desired output field $\mathbf{E}_{\text{out}}^{i}$.

**In the following we focus on the Green dyadic method [10] in order to find an expression for the matrix $\mathbb{X}$, but any other solver of Maxwell's equations could also be used, equivalently. We look first for the self-consistent electric field $\mathbf{E}_{\text{out}}$ inside the nano-object by discretizing its volume $V$ into $N$ identical sub-wavelength meshes located at the positions $\mathbf{r}_i$, where $i \in [1, N]$. This procedure generates a system of $N$ vectorial equations with $N$ unknown fields $\mathbf{E}_{\text{out}}(\mathbf{r}_i)$:**

$$\mathbf{E}_{\text{out}}(\mathbf{r}_i) = \mathbf{E}_{\text{in}}(\mathbf{r}_i) + \chi \frac{V}{N} \sum_{j=1}^{N} \mathbb{S}(\mathbf{r}_i, \mathbf{r}_j) \mathbf{E}_{\text{out}}(\mathbf{r}_j), \qquad (3)$$

**where $\chi = \dfrac{\varepsilon_{ob}(\omega) - \varepsilon}{4\pi}$, $\varepsilon_{ob}(\omega)$ and $\varepsilon$ are the frequency dependent dielectric functions of the metal and of the surroundings, and $\mathbb{S}(\mathbf{r}_i, \mathbf{r}_j) = \mathbb{S}_0(\mathbf{r}_i, \mathbf{r}_j) + \mathbb{S}_{surf}(\mathbf{r}_i, \mathbf{r}_j)$ is the Green dyadic** tensor including a free-space term $\mathbb{S}_0(\mathbf{r}_i,\mathbf{r}_j)$ and a reflection term $\mathbb{S}_{surf}(\mathbf{r}_i,\mathbf{r}_j)$ accounting for the presence of a surface [10]. Eq. (3) forms a linear system that can be rewritten in a simplified notation, after removing the self-consistency [10]:

$$\mathbf{E}_{in} = \mathbb{M}\mathbf{E}_{out}, \qquad (4)$$

where $\mathbf{E}_{in}$ and $\mathbf{E}_{out}$ are $3N$-supervectors defined by $\mathbf{E}_{in} = (\mathbf{E}_{in}(\mathbf{r}_1), \mathbf{E}_{in}(\mathbf{r}_2), ..., \mathbf{E}_{in}(\mathbf{r}_N))$ and $\mathbf{E}_{out} = (\mathbf{E}_{out}(\mathbf{r}_1), \mathbf{E}_{out}(\mathbf{r}_2), ..., \mathbf{E}_{out}(\mathbf{r}_N))$. $\mathbb{M}$ is an invertible ($3N \times 3N$) matrix [10], formed by the ($3 \times 3$) elements $\mathbb{M}_{i,j} = \delta_{i,j} - \chi \frac{V}{N} \mathbb{S}(\mathbf{r}_i, \mathbf{r}_j)$.

As a test structure to illustrate the method, **we consider a V-shaped ensemble of five 80x80x40-nm gold pads (left inset of Fig. 2 (a)) lying onto a glass substrate ($\varepsilon_{ob}(\omega)$ is taken from [11], $\varepsilon_{glass} = 2.3$ and $\varepsilon_{air} = 1$).** In Fig. 2 (a), the calculated far-field spectra are plotted for a plane wave illumination linearly polarized along the x- and y-direction. Upon each polarization, the V-structure features two resonances each associated with a different electric near field intensity distribution (right insets of Fig. 2). While changing the direction of the linear incident polarization and/or the wavelength leads to significant changes in the near field response of the nanosystem, the degree of control that can be exerted remains strongly limited by the fixed geometry.

Fig. 3 (a-c) shows one scenario of sub-λ control that one would like to achieve over the V-shaped nanostructure **in which light is selectively coupled to the dipolar mode (resonance at around 560 nm) of each of the adjacent pads of the ensemble**. In practice, the desired electric field $\mathbf{E}_{out}^i$ is determined by computing the response of a single isolated pad **upon a focused Gaussian illumination. This field is then imposed to the target pad of the complex nano-system and used as input for the DOPTI algorithm. In our simulation, the focusing optics was modeled through a 1.25 numerical aperture (NA) aplanatic lens with a 3-mm entrance radius [12].** The distance between the centers of two adjacent pads is less than half the limit of diffraction (approximately 270 nm at a wavelength of 560 nm with 1.25 NA), represented by the pink circle in Fig. 3.

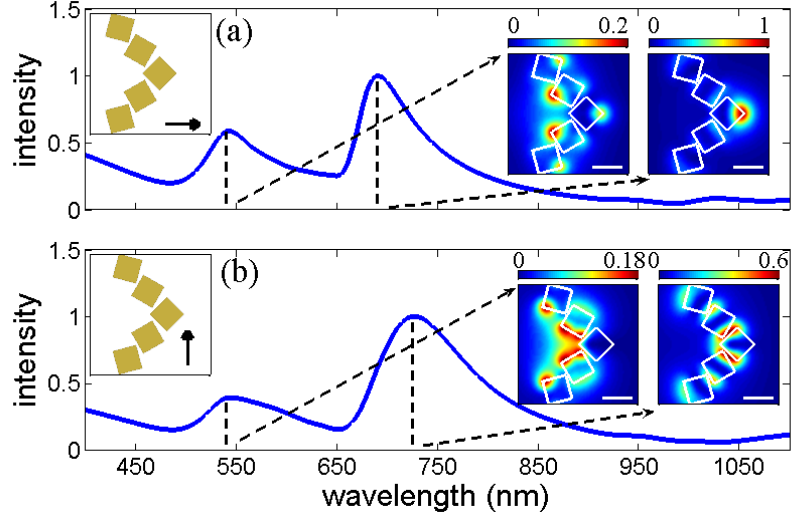

FIG. 2: (color online). Partial control of the optical response of a V-shaped gold structure upon plane wave illumination. (a) Normalized far-field scattering spectrum of a V-shaped nanostructure (in the left inset) illuminated by a plane wave linearly polarized along the x-direction. The right insets represent the normalized electric near-field intensities at 545 nm and 690 nm (scale bar = 100 nm). (b) Normalized far-field scattering spectrum of the same nanostructure illuminated by a plane wave linearly polarized along the y-direction (in the left inset). The right insets represent the normalized electric near-field intensities at 550 nm and 725 nm (scale bar = 100 nm). **All the near-field intensities were calculated 20 nm above the pads [10].**

Eq. (4) always provides a mathematical solution for the incident field $\mathbf{E}_{in}^{i}$ associated to any distribution of the desired field $\mathbf{E}_{out}^{i}$. However, the practical implementation of this inversion approach depends on whether there is a physical incident field close enough to the ideal mathematical solution that can be created from the Gaussian beam of a laser. **In order to ensure a physical solution, we impose that the field $\mathbf{E}_{in}^{i}$ must be decomposable on a basis $\mathbb{H}$ of *n*** focused Hermite-Gaussian beams, whose symmetry axis coincides with the origin of the Cartesian coordinates $\mathbf{r}_0 = \mathbf{0}$. Our choice fell on this basis in order to establish a direct link with standard experimental conditions, where the incident field is the result of focusing a paraxial beam through an aplanatic lens [12]. One the one hand, the HGs form an orthogonal basis that can completely describe any paraxial beam before the lens [13-14] and, on the other hand, can feature, after focusing, phase singularities at the sub-wavelength scale [15]. $\mathbf{E}_{in}^{i}$ can therefore be expressed as:

$$\mathbf{E}_{in}^{i} = \mathbb{H}\boldsymbol{\beta}, \qquad (5)$$

where $\mathbb{H}$ is a **(3*N* x *n*)** matrix whose columns are the elements of the basis and $\boldsymbol{\beta}$ is the vector containing the complex coefficient of the superposition. **In this work, the expansion was truncated to the first 7 orders to limit the phase complexity so that $\mathbb{H}$ becomes a (3*N* x 14) matrix (for each order, both orthogonal polarizations along the Cartesian axis, *x* and *y*, are considered)**.

**Inserting Eq. (5) into Eq. (4) enables us to make clear the linear relationship between the unknown coefficients of the superposition β and $\mathbf{E}_{out}^{i}$ :**

$$\mathbf{E}_{out}^{i} = \mathbb{M}^{-1}\mathbb{H}\boldsymbol{\beta}. \qquad (6)$$

**Eq. (6) is, therefore, a linear over-determined system with 3*N* equations and *n* unknowns (3N>>n). As a consequence, depending on the desired near field pattern, a unique solution for Eq. (6) does not always exist; a turnaround approach is to look for the value $\boldsymbol{\beta}^{a}$ of the coefficients of the superposition that offers the best physical approximation $\mathbf{E}_{out}^{a}$ to our ideal electric field $\mathbf{E}_{out}^{i}$, such that:**

$$\mathbf{E}_{out}^{i} = \mathbf{E}_{out}^{a} + \mathbf{r} = \mathbb{M}^{-1}\mathbb{H}\boldsymbol{\beta}^{a} + \mathbf{r}, \qquad (7)$$

where the 3$N$-supervector **r** is the vector of residuals or error vector, that only vanishes when Eq. (6) has a unique solution. This becomes a standard fitting problem that can be solved by adopting any linear algorithm, such as a linear Least Mean Squares algorithm (LMS) [9]. The LMS solution is, for example, given by [9]:

$$\boldsymbol{\beta}^a = \left(\mathbb{X}^h \mathbb{X}\right)^{-1} \mathbb{X}^h \mathbf{E}^i_{out}, \quad \mathbb{X} = \mathbb{M}^{-1}\mathbb{H}, \tag{8}$$

$$\mathbf{r} = \mathbf{E}^i_{out} - \mathbf{E}^a_{out} = \mathbf{E}^i_{out} - \mathbb{X}\boldsymbol{\beta}^a, \tag{9}$$

where the matrix $\mathbb{X}^h$ is the transpose conjugate (hermitian) of $\mathbb{X}$. The LMS solution also gives direct insight on how physical and viable our initial ideal field $\mathbf{E}^i_{out}$ is in terms of relative error $r$ and absolute efficiency $\eta$, as defined in point C of the supplementary information. Overall, the error $r$ and efficiency $\eta$ quantify the fidelity which our approximated solution describes the intended output field with, and the coupling efficiency of the desired incident field, respectively. According to our definition of the efficiency, for $\mathbf{E}^a_{out}$ to reach the same maximum intensity as $\mathbf{E}^i_{out}$, the total power of the incident field has to be exactly $\eta$ times higher. Note that for most applications, our restrictive definition for the error $r$ provides a worst-case scenario. For instance, a definition of the error based on simply concentrating light around a single gold pad (i.e. without necessarily exactly reproducing its dipolar mode pattern) would lead to much smaller errors of those in Fig. 3.

Fig. 3 (d-f) show the reconstructed near-field maps of $\left\|\mathbf{E}^a_{out}\right\|^2$ associated to the desired near field patterns of Fig. 3 (a-c). The corresponding desired and reconstructed phase distributions are presented in the supplementary material (Fig. S2). Fig. 3 (g-i) plot the associated magnitude and phase of the coefficients of the vector $\boldsymbol{\beta}^a$. The error $r$ is found to be of respectively 56%, 20% and 11%, while the efficiency is respectively 13, 6.5 and 7.8. At this stage, it becomes clear that an approach exclusively based on reproducing the required incident intensity pattern (without considering the phase distribution), for instance through a superposition of diffracted limited Gaussian spots, would suffer from a much lower level of control.

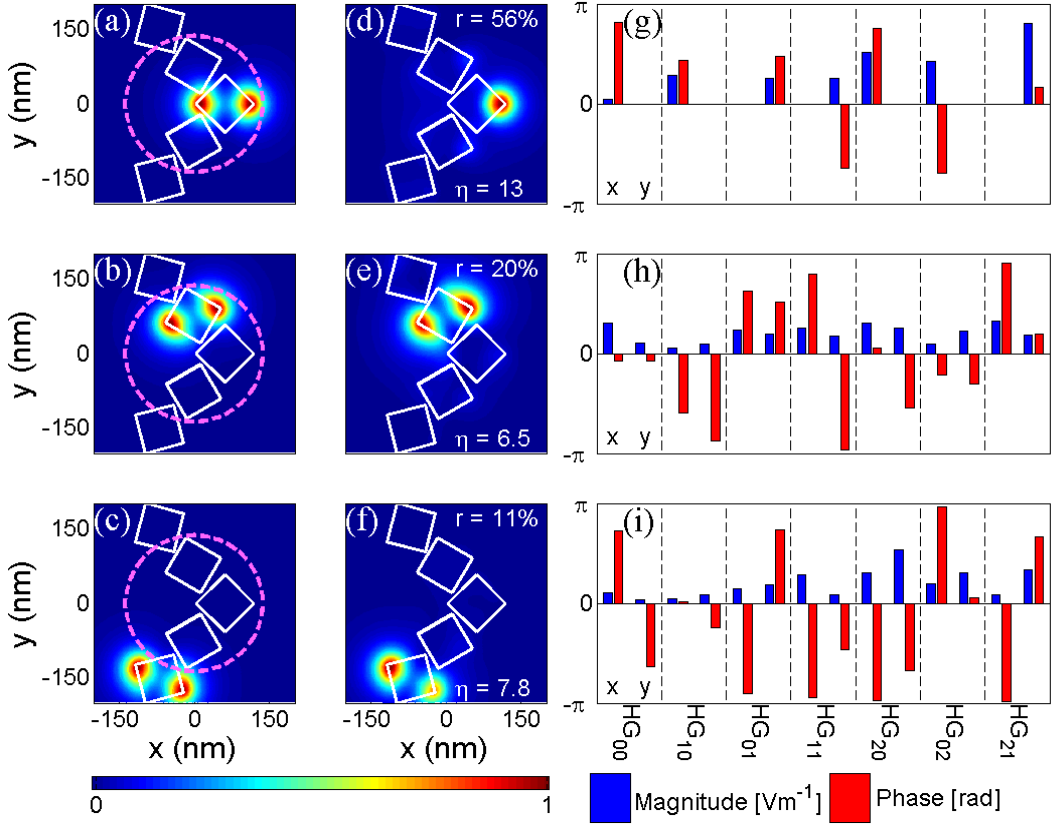

FIG. 3: (color online). DOPTI sub-λ control in a V-shape structure. (a-c) Normalized electric near-field intensity distribution for three cases of desired fine control at 560 nm. The pink dashed circle depicts the diffraction limit. (d-f) Normalized reconstructed electric near-field intensities calculated from the three intended output near field patterns in (a-c). Errors and efficiencies are also reported. **All the near-field intensities were calculated 20 nm above the pads [10]**. (g-i) Corresponding magnitude (blue) and phase (red) of the $\beta$ coefficients of the superposition on a base of 14 HG beams. Every order of the base of HG beams appears with the two orthogonal polarization along x and y. The values of both magnitude and phase are expressed on the same scale, between -π and π.

As a step forward, we apply the DOPTI protocol to the design of a nano-sized structure formed by a cross-shaped array of **five 80x80x40-nm gold pads**, as shown in Fig. 4 (a). We aim here at employing this nanostructure to selectively optically address several nearby nano-objects, such as nanoemitters, located at positions A, B and C. **In this case, the input of the DOPTI algorithm is the electric field $E_{out}^i$ derived by concentrating light on the left corner of a single pad of the ensemble.** Tests on the cross-like nanostructure were performed at two wavelengths, 560 nm and 690 nm, whose limits of diffraction sits approximately at 270 nm and 335 nm (represented by the two circles in Fig. 4 (a)). For both wavelengths, the separation distance between points A, B and C is substantially less than half the limit of diffraction. Fig. 4 (b) and (c) show how it is possible to selectively localize light at point A, B or C for both wavelengths with intensity ratio up to 1/100. The corresponding normalized electric near-field intensities are reported in Fig. 4 (d), (e) and (f) for 560 nm wavelength, and in Fig. 4 (g), (h) and (i) for 690 nm together with the respective errors and efficiency, **as defined in point C of the supplementary information**. Accurate control of the field concentration at the three points is successively achieved at both wavelengths, showing that **the DOPTI method is almost independent** from the illumination wavelength. Nevertheless, we observe that efficiency and errors are better at 560nm than at 690nm, most probably because at this wavelength the intended field patterns are closer to the eigenmodes of the structure upon plane wave illumination.

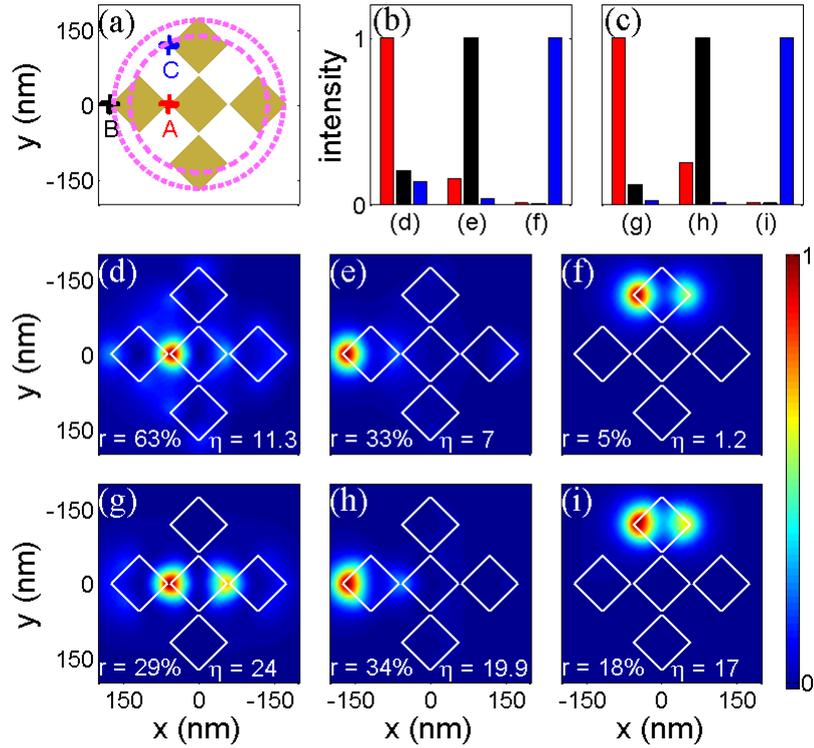

FIG. 4: (color online). Design of a nano-optical element able to selectively address several nearby nano-objects. (a) Cross-like structure, the two violet dashed circles represent the limit of diffraction for 560 nm and for 690 nm respectively. (b-c) Intensity localization in points A (red), B (black) and C (blue) at 560nm and 690 nm respectively. (d-f) Normalized electric near-field distribution calculated for the three cases in (b) at 560 nm. (g-i) Normalized electric near-field distribution calculated for the three cases in (c) at 690 nm. **All the near-field intensities were calculated 20 nm above the pads [10].**

As a conclusion, we have described and tested a universal protocol that makes it possible to determine the physical incident light field required to achieve a desired near field response of a nanostructure. Even though there is no practical limitation on the choice of the desired near-field pattern, smaller errors and higher efficiency are expected for patterns that are close to a superposition of the eigenstates of either the whole structure or a portion of it. This protocol combined with spatial phase shaping by a Spatial Light Modulator (SLM) offers a realistic approach towards the dynamical control of the near field of any linear nanosystems.

This work was supported by the Spanish Ministry of Sciences through Grants TEC2007-60186/MIC and CSD2007-046-NanoLight.es and La Fundació CELLEX Barcelona. The authors would like to thank C. Girard, L. Novotny, G. Volpe, G. Baffou and J. Renger for some fruitful discussions.